\documentstyle[prb,aps,preprint]{revtex}
\begin{document}

\begin{title}
{ 
Atomic and electronic structure of ideal and reconstructed
 $\alpha$-Sn (111) surface
}
\end{title}

\author{Zhong-Yi Lu$^{a,b}$, Guido L. Chiarotti$^{a,b}$, 
S. Scandolo$^{a,b}$, 
and E. Tosatti$^{a,b,c}$}
\address{
$^{a)}$ International School for Advanced Studies (SISSA),
Via Beirut 2, I-34014 Trieste, Italy\\
$^{b)}$ Istituto Nazionale di Fisica della Materia (INFM), 
Trieste, (Italy)\\
$^{c)}$ International Center for Theoretical Physics (ICTP), 
P.O. Box 586, I-34014
Trieste, Italy
}

\date{\today}
\maketitle

\begin{abstract}
We have carried out an {\em ab-initio} study 
of $\alpha$-Sn (111), with 
the aim of predicting and understanding its structure, 
reconstructions, and electronic states. 
We consider a variety of structural possibilities,
and optimize them by moving atoms 
according to Hellmann-Feynman forces.
Our results indicate 
that the unreconstructed surface is highly unstable,
while a variety of reconstructions 
compete for the true ground state.
Extrapolated trends 
from diamond to Si to Ge are well borne out, with
a $2\times 1~\pi$-bonded chain reconstruction 
prevailing in the absence of adatoms,
and a $c(4\times 2)$ or $(2\times 2)$ basic adatom-restatom unit 
reconstruction otherwise. Accompanying 
surface bucklings are in both cases larger than in Si and Ge,
with consequently large ionic charge transfers predicted.
Search for a $\beta$-Sn-like metallic 
state of the surface turned out 
to be inconclusive.

\end{abstract}



\section{Introduction}

Essentially all semiconductor surfaces are reconstructed. 
As is well known, this can be traced 
back to the strong covalency of their bulk band structure.
In fact, the uniqueness of covalent systems, 
as opposed to, for example, regular metals, 
is that bulk covalent bonding is very directional and relatively
inflexible. Consequently, it generally costs 
too much energy to locally rearrange the electron wave 
functions at the surface 
(as required by the presence of broken bonds), 
without some major lattice readjustments. 

In the group-IV insulators and semiconductors, 
known to crystallize in the diamond 
structure, the best studied face by far is 
the (111), where well documented results are
available for diamond, Si and Ge. 
Both experiment and theory indicate that a variety of 
different reconstruction mechanisms are realized. 
A summary is presented in
Table \ref{trend}. One can see that all (111) surfaces possess a 
$\pi$-bonded chain $2 \times 1$ reconstructed state. 
As a true ground state,
however, $\pi$-bonded chains only prevail in diamond,
possibly reflecting the difficulty of obtaining well annealed
adatoms on this surface. On Si and Ge (111), the adatom
reconstructions, $(7\times 7)$ [Ref.\onlinecite{7x7}] 
and $c(2\times 8)$ [Ref.\onlinecite{c2x8}] respectively, 
have lower energy,
whereas the $2\times 1$ $\pi$-bonded chain reconstruction remains 
well-defined, but energetically metastable.

In all cases: 
diamond, Si and Ge (111), the metallic character of 
the ideal, symmetrical $\pi$-bonded chain 
reconstruction\cite{Pandey} is 
removed by some accompanying distortion. 
This distortion can consist either of a slight 
dimerization of the chain, as is probably the 
case in diamond\cite{Galli_c2x1}, 
or of a vertical buckling, 
as in Si\cite{Northrup,Ancilotto_si2x1,LB} 
and Ge \cite{Takeuchi_ge2x1}. Although experimental 
buckling data for diamond and Ge (111) are not yet available, 
{\em ab-initio} calculations 
exist for these geometrical distortions.  
The size of the predicted 
chain bucklings exhibits an interesting 
trend, namely it is zero in diamond\cite{Galli_c2x1}, 
about 0.4 \AA\ in silicon\cite{Ancilotto_si2x1}
(where experimental data\cite{LB} confirm this magnitude),
and 0.8 \AA\ in germanium \cite{Takeuchi_ge2x1}.
The absence of buckling 
in carbon has been ascribed to the larger 
intra-atomic Coulomb repulsion, which prevents the ionic 
charge transfer implied by the buckling\cite{Galli_c2x1}. 
The decrease of Coulomb repulsion with increasing atomic 
number is consistent with the trend towards a larger buckling.

When adatoms are present, as in annealed Si and Ge (111), 
then an adatom-rest atom reconstruction prevails. 
Interestingly, the adatom reconstruction again 
implies surface ionicity \cite{Meade_sige2x2},
with an electron charge transfer 
between the adatom and the restatom. 

Besides $\pi$-bonded chains and 
adatom-restatom reconstructed states, 
a third kind of state, namely the displacive 
$2\times 1$ buckled reconstruction, originally proposed
by Haneman\cite{Haneman}, may also play a role. Although never 
found to be stable experimentally, and never stable 
theoretically either, it does make a brief appearance
as a metastable transition state 
in the transformation path from 
ideal $1\times 1$ to $2\times 1$ $\pi$-bonded chain, 
in Si \cite{Ancilotto_si2x1} and Ge (111) \cite{Takeuchi_ge2x1}, 
as found by {\em ab-initio} molecular dynamics. 
This does not seem to occur on C (111) \cite{Galli_c2x1}, 
probably again because the Haneman state is 
ionic, and thus very costly in carbon. 

Finally, the last entry in Table \ref{trend} 
considers, as a fourth type of reconstruction,
the possible transformation of a thin surface layer from 
semiconducting to metallic. 
This possibility is suggested by the bulk
phase diagrams of Si, Ge and Sn, 
where fully metallic phases completely 
surround the semiconducting phases, under any combination of 
either high pressure or temperature. 
There is considerable evidence that surface metallization does in 
fact play a relevant role 
at high temperatures\cite{Takeuchi_metal}.
The finite temperature 
behavior is however outside the scope of the present work.
Although this transformation (as predicted, for example, 
in Ga\cite{Marco_gallio}) 
does not appear to prevail at T=0 K in Si
and Ge, it could still take place in $\alpha$-Sn, 
which is closer to being metallic. 

Throughout the literature, only the (100) surface of 
$\alpha$-Sn appears to have been considered so far\cite{Yuan}. 
This paper is devoted to what is therefore 
a first study of the $\alpha$-Sn (111) surface. 
In particular, since there are no experimental data available, 
{\em ab-initio} calculations are required 
to compare the various possible 
reconstructions, and understand the trends relative to the other 
semiconductors just described. By extrapolating 
the trends of Table \ref{trend} 
we are led to expect the following:\\ 
i) the $2\times 1$ $\pi$-bonded chain 
reconstruction should be present 
also in $\alpha$-Sn(111), 
with an even more pronounced buckling than in Ge;\\ 
ii) This reconstruction will be in energetic competition 
with some adatom/restatom reconstructions, 
the latter being likely to prevail;\\ 
iii) A surface metallization mechanism 
is also possible because of the presence
of the bulk metallic $\beta$-Sn phase, 
energetically very close to the $T=0$ bulk ground state
$\alpha$-Sn.  

In this work we shall find that (apart from metallization, which
turns out to be a harder problem) these trends are generally 
well borne out by {\em ab-initio} calculations. 
An adatom-restatom reconstruction is 
predicted to be the annealed ground state
of $\alpha$-Sn (111). 
Although the exact periodicity cannot be predicted,
our geometry and energetics based on an elementary 
$c(4\times 2)$ or $2\times 2$ unit should be fairly reliable.

This paper is organized as follows. 
First of all, we calculate electronic structure, 
total energy and optimal geometry of bulk Sn, both in the $\alpha$
semimetallic phase, 
and in the $\beta$ metallic phase. This is described
in Section II. Next, in section III 
we use the same ingredients for a 
surface calculation in the usual slab geometry. 
We deal first with the ideal,
unreconstructed $\alpha$-Sn (111) surface, 
whose properties we study
without and with surface relaxation. 
We find partly filled dangling
bond states in the ideal surface electronic structure, 
which as usual suggest that
this is not a stable surface. 
This is confirmed in Section IV, where
a $2\times 1$ displacive reconstruction is shown to lower
surface energy without energy barriers. 
The more interesting $\pi$-bonded 
chain reconstruction is considered in Section V. It is found
that a very strongly buckled $2\times 1$ $\pi$-bonded chain state 
exists for $\alpha$-Sn (111), 
its energy being better than that of the 
Haneman state. Section VI is devoted to 
the alternative possibility of adatom/restatom reconstructions. 
We find this latter possibility to be energeticall ythe best, 
particularly in the $c(4\times 2)$ local geometry. Lastly,
in Section VII we briefly describe an attempt at finding a stable
metallic reconstructed surface ground state, which however turns
out to be inconclusive. 
Section VIII summarizes the main conclusions
of this work on $\alpha$-Sn (111), 
confirming the trends extrapolated 
from diamond, Si and Ge, and stressing our predictions for
future experimental work.

\section{Bulk properties}

The zero temperature structure of bulk Sn is the $\alpha$ phase
with a diamond lattice \cite{Donohue} of spacing 
6.483 \AA\ (at 90 K) \cite{Rice}. 
As is well known, if the temperature 
is raised to $286$ K at atmospheric pressure, 
semimetallic $\alpha$-Sn 
transforms into the fully 
metallic $\beta$-Sn phase\cite{Jayaraman}. 
Under pressure, this transformation occurs 
at lower temperatures, extrapolating to T=0 K
at a pressure of 5 Kbar \cite{Jayaraman,Cohen}.
Total energy LDA (local density approximation) 
calculations \cite{Cohen} 
confirm that at zero temperature and zero pressure 
the $\beta$ phase is energetically 
very close to the $\alpha$ phase, 
their energy difference being 44 meV/atom,
with a predicted transition pressure to the $\beta$ phase of 8 Kbar.
The relative success of LDA calculations 
in predicting very closely these extremely
delicate differences provides confidence 
in extending this approach 
towards exploring surface properties.

In the present {\em ab-initio} calculations 
for $\alpha$-Sn (bulk and surface)
and $\beta$-Sn (bulk) we model the electron-ion interaction with
a norm-conserving pseudopotential \cite{Stumpf} in the 
Kleinman-Bylander form \cite{KB} with $s$ and $p$ nonlocality, 
and use the standard LDA for the electron-electron interaction. 
We expand the Kohn-Sham orbitals in plane waves with an energy 
cutoff of 12 Ry in all 
calculations (bulk and surface) for consistency. For the 
$\alpha$-Sn bulk calculation, 10 special k-points\cite{Chadi} 
are sampled in the Irreducible Brillouin Zone (IBZ) 
of the diamond phase.
For the metallic $\beta$ phase, instead, we find that 
160 k-points are needed to describe 
accurately the IBZ summations, with
an additional gaussian spreading of 0.14 eV \cite{spreading}. 
We find a lattice constant $a=6.446$ \AA\ 
(exp. $a=6.483$ \AA\ \cite{Rice}) 
for $\alpha$-Sn  and $a=5.768$ \AA\ 
(exp. $a=5.812$ \AA\ \cite{Rayne}) 
for $\beta$-Sn (for simplicity, we fixed the c/a ratio
to the experimental value of 0.546). 
We also find that the $\beta$-phase is
disfavored with respect to the $\alpha$-phase by 17 meV/atom, 
with an $\alpha \to \beta$ transition pressure of 5 Kbar 
in excellent agreement with experiment.
We have tested that increasing the energy cutoff to 35 Ry
and the k-point integration to 200 points 
in the IBZ of the $\beta$-phase
does not alter significantly these results. On the contrary, 
if the energy cutoff is less than 12 Ry, e.g. 10 Ry, 
the transition pressure 
becomes 2.5 Kbar and the energy difference drops to 9 meV/atom.

The electronic band structure of $\alpha$-Sn is reported
in Fig. \ref{bulk_band} and that of $\beta$-Sn\cite{band_beta} 
in Fig. \ref{beta_band}. 
As can be seen, in particular, the semimetallic character 
of the $\alpha$-phase (zero gap at $\Gamma$) and the fully metallic
character of the $\beta$-phase 
are of course correctly borne out in our calculations.

\section{The relaxed, unreconstructed $\alpha$-Sn (111) surface}

We study the surface properties of $\alpha$-Sn (111) by a standard 
slab calculation. One of the two surfaces of the slab is frozen 
in its ideal geometry, together with the 
first three adjacent layers (a total of two rigid bilayers).
The atoms belonging to the remaining layers 
are allowed to fully relax guided  
by the corresponding Hellmann-Feynman forces. Convergence is 
assumed when forces are less than 4 meV/\AA\ . 
The number of atomic layers in the slab 
is fixed to a total of 12 layers 
(not including the adatom layer when present). 
Each layer consists of either one, two, or four Sn atoms, 
in correspondence to 
choosing $1\times 1,$ $2\times 1,$ and $2\times 2$
[or c($4\times 2$)] surface cells, respectively. The number
of ``vacuum layers'' is 
fixed to 6 (vacuum thickness $\sim 11$ \AA\ ), 
and the number of k points in the 
Irreducible Surface BZ (ISBZ) is 
chosen according to the size of the surface 
cell and its geometry, as described case by case. 
The initial atomic positions 
are chosen according to the 
calculated equilibrium bulk lattice spacing
($a_\circ = 6.446$ \AA\ ).
This guarantees that the forces on the atoms of the central
layers are smaller than 4 meV/\AA\ for all the surfaces studied. 
In all cases, surface energies $E_{surf}$ 
are given as $E_{surf} = E_{slab} - N E_{bulk}-E_{surf}^{ideal}$, 
where $E_{slab}$ is the total energy of the slab, 
$E_{bulk}$ is the energy per atom
of bulk $\alpha$-Sn as computed 
in section II [$E_{bulk}$=-96.753 eV/atom] above, 
$N$ is the total number of atoms in the slab. Furthermore 
$E_{surf}^{ideal}=[E_{slab}^{ideal}-NE_{bulk}]/2$ is the energy of
the frozen ideal surface. Here $E_{slab}^{ideal}$ is
the total energy of a slab with both surfaces rigid and ideal,
computed using the same supercell 
geometry and k-point set as $E_{slab}.$ 
We thereby have to repeat ideal surface calculation several 
times in different supercells and k-point sets corresponding to
different surface reconstruction slab calculations in order to
compare different reconstructed surface energies. 
The convergence of our results with respect to the k-points 
sampling has been tested 
for each surface by increasing the k-point number 
after the atomic relaxation. Accordingly, we can estimate 
our overall energy resolution to be about 10 meV/$(1\times 1)cell.$ 
We also constructed the electronic band 
structure of the relaxed surfaces, by using the 
Kohn-Sham eigenvalues of a 10-layer slab, obtained by removing
the seven bottom layers of the simulation slab 
and replacing them by the inverted
image of the topmost five relaxed layers. 
The reason for this procedure is threefold. First, we get rid 
of the undesired states related to the bottom rigid ideal surface. 
Second, we increase the effective symmetry of the supercell, thus 
decreasing the computational effort 
of the band structure calculation. 
Third, although the interaction of identical surface 
states belonging to opposite surfaces 
generally lifts their degeneracy
(again an undesired effect), 
their average energy still corresponds, to first order, to the 
noninteracting value in the ideal case of an infinite slab. 
 
As a first case we have considered 
the $(1\times 1)$ unreconstructed surface. Six special 
k-points in the hexagonal ISBZ have been used \cite{kpoints}.  
The electronic band structure 
of the ideal $(1\times 1)$ surface is reported in
Fig. \ref{ideal_band}. There are various surface 
states lying in the projected gaps. 
The surface states crossing the Fermi level inside the 
fundamental gap are clearly related to 
the presence of unsaturated surface dangling bonds.
This feature is common to all the other group-IV 
insulator and semiconductors, and is responsible for
the high grade of instability of the ideal (111) surface, 
as mentioned. The electron density corresponding 
to the surface state is analysed
in Fig. \ref{ideal_charge}, 
and reveals a high degree of surface localization
as well as a clear dangling-bond character. 

As the next step, we allowed the surface, i.e.
all atoms in the eight topmost surface layers, 
to relax, according to
Hellmann-Feynman forces, so as to reduce surface energy. 
Despite the presence
of unsaturated dangling bonds, 
the result of this energy minimization shows that surface atoms 
do not relax significantly (see Tabel \ref{energy}), with a 
surface energy gain of only 7 meV/$(1\times 1)cell$, and
a downward relaxation of the top layer of 0.02 \AA\  .
We also checked that increasing the k-point number
from six to eighteen changed none of the above results.

\section{Haneman ($2\times 1$) ``buckled atom'' reconstruction}

In the previous calculation, 
reconstruction was forbidden by symmetry.
If the symmetry constraints 
imposed by the choice of a $(1\times 1)$ 
cell are relaxed, the ideal surface is provided with a simple
mechanism for the partial 
saturation of the dangling bonds. One such 
mechanism, first proposed by Haneman \cite{Haneman}, 
consists of a simple in-out 
buckling of the topmost layer, resulting 
in a $(2\times 1)$ displacive reconstruction. The inward motion 
of one surface atom implies 
an $sp^3\rightarrow sp^2$ rehybridization,
and a more $p_z$-like dangling bond for that atom. 
The outward motion of the other atom,
by contrast, causes dehybridization, 
and a more $s$-like dangling bond.
Since in the atom $E_s<<E_{p_z},$ 
electrons will flow from the inward 
to the outward atom. This charge transfer empties and saturates 
respectively the two dangling bonds and 
therefore stabilizes the surface.
This $(2\times 1)$ reconstruction, 
although actually never observed,
has been recently suggested 
to play the role of a transition state,
a kind of ``stepping stone'' in the dynamical process leading 
from an unreconstructed state towards, e.g., a $(2\times 1)$ 
$\pi$-bonded chain state\cite{Takeuchi_ge2x1}.
In order to study the possible occurence of such a buckled atom
reconstruction on $\alpha$-Sn (111),    
we have repeated our calculation in a larger $(2\times 1)$ 
surface cell, using 4 special k-points 
in the rectangular $(2\times 1)$ ISBZ\cite{kpoints}. 
We find, indeed, that the surface spontaneously buckles as in
the Haneman distortion, against which the ideal surface is 
therefore unstable, 
very much as Ge (111) does\cite{Takeuchi_ge2x1}.  
The energy gain from the ideal to the optimal buckled geometry 
(Table \ref{energy}), is about 0.22 eV/$(1\times 1)cell,$ 
measured relative to the energy of the ideal
surface computed using the same $(2\times 1)$ 
supercell and k-point set.
The buckling of the top layer 
is found to be enormous, namely 1.23 \AA\ . 
The final atomic coordinates of the relaxed top five atomic 
layers are given in Table \ref{buck_pos}.

\section{$(2 \times 1)$ $\pi$-bonded reconstruction}

The instability of the ideal (111) surface 
against $(2\times 1)$ buckling is
interesting, but probably academic, except possibly in dynamics.
In our pursuit of the true (111) ground state surface structure,
we consider next the surface reconstruction geometries which are
experimentally observed in the other group-IV 
insulator and semiconductors. Here we consider 
the (111) $2\times 1$ $\pi$-bonded chain reconstruction. 
In the lack of any experimental and theoretical data,
we arranged the surface atoms in the structure
proposed by Pandey \cite{Pandey}. Pandey's 
structure can be obtained 
by simultaneously depressing one surface
atom into the second layer and 
correspondingly raising one second layer atom,
so that i) the number of dangling bonds per surface atom is
the same as before; ii) all bond lengths are 
set to the bulk value.
The atomic coordinates 
in Pandey's structure are given in Table \ref{pi_pos}. 
Subsequently, we allowed the atomic positions 
to relax according to their {\em ab-initio} forces, 
using the same supercell and same k-point sampling as 
in the previous section. At convergence, the energy gain of this
$2\times 1$ reconstruction, with respect to the ideal surface, is
0.24 eV/$(1\times 1)cell$ (Table \ref{energy}), 
i.e., larger than that of the previous section.   
The optimal atomic coordinates of the top five atomic 
layers are given in Table \ref{pi_pos}. The bond lengths 
within the $\pi$-bonded chain are 2.80 \AA\  , only 0.4 \% longer 
with respect to the ideal bulk bond length (2.79 \AA\ ). 
The chain buckling (1.15 \AA\ ) is however very large if compared
with that of Si or Ge (0.4 \AA\ \cite{Ancilotto_si2x1,LB}, and 
$0.8$ \AA\ \cite{Takeuchi_ge2x1}, respectively), 
fully confirming the 
trends discussed in Section I and listed in Table \ref{trend}. 
The magnitude of this buckling is so large, 
that the whole $\pi$-bonded 
chain is now lying onto an essentially 
vertical plane (see Fig. \ref{2x1_charge}).
The large chain buckling is accompanied by 
a large electron transfer from the lowered to the raised 
chain atom,  which largely saturates
the raised atom dangling bond. This is clearly seen in Fig. 
\ref{2x1_charge}, 
where the charge distributions of the highest occupied 
(panel (a)) and lowest unoccupied states (panel (b)) are shown. 
This charge rearrangement is 
accompanied by the opening of a large gap
in the dangling bond surface state, as seen in the electronic
structure of Fig. \ref{2x1_band}. 
Here the highest occupied (lowest unoccupied) surface states 
correspond to raised (lowered) atom dangling bonds respectively.
Finally, since the alternative possibility of a dimerization of 
the chain was excluded in our cell 
due to symmetry constraints, we tried slightly  
dimerizing the initial, unbuckled $\pi$-bonded chain, 
and also the final, 
fully buckled one (thereby doubling the number 
of k-points in the ISBZ). 
However, we found these configurations 
to be energetically disfavored 
with respect to the undimerized chain.  
 
\section{Adatom / restatom reconstructions}

As the next likely candidate 
for the reconstruction of $\alpha$-Sn (111),
we now consider the adatom / restatom reconstruction, known to 
be the stable mechanism for both Si and Ge. Although this 
reconstruction shows up with rather more 
complex surface unit cells, such as the 
$(7\times 7)$ DAS model in Si\cite{7x7} 
and the $c(2\times 8)$ in Ge\cite{c2x8}, 
the building block of this class of reconstructions is simple. 
It is based on the presence of one adatom every four 
$(1\times 1)$ first-layer atoms. 
The adatom sits in a threefold site, 
saturating three first-layer atoms, 
leaving one (the restatom) unsaturated.
Of the two available threefold sites, namely 
$T_4$ (on top of a second-layer atom), 
and $H_3$ (hollow site), the 
adatoms prefer, at least 
in Si and Ge\cite{Meade_sige2x2,Takeuchi_ge2x2},
the $T_4$ site. 
Moreover, the $T_4$ site adatoms may still be arranged 
in a $(2\times 2)$ or $c(4\times 2)$ geometry. 
For instance in Ge (111) $c(2\times 8)$, they are stacked in 
alternated $(2\times 2)$ and $c(4\times 2)$ cells.
In this work we assume the pure $(2\times 2)$ [or pure
$c(4\times 2)$] $T_4$ structure as the prototype 
adatom / restatom reconstruction, 
restricting for simplicity our analysis to 
this case only. 
As for the $\pi$-bonded reconstruction case, we have no
data to guide us. 
We start with an ideal $T_4$ position for the adatom (see
Table \ref{2x2_pos}) such that the 
lengths of the adatom bonds with the three first-layer atoms
are equal to the bulk bond-length. 
The relative positions of atoms are initially chosen 
according to the $(2\times 2)$ [or $c(4\times 2)$] 
case of Table \ref{2x2_pos} [or Tabel \ref{c2x4_pos}].
The $k$-point sampling of this larger supercell is 
restricted to a single special point 
(mean-value point) of the 
hexagonal $(2\times 2)$ ISBZ\cite{kpoints}. 
We then calculate Hellmann-Feynman forces, and
let the atoms relax to the equilibrium positions. We also find 
that increasing the number of 
special k-points of the relaxed surface
to three and six does not change 
the surface energy within 5 meV/$(1\times 1)cell$. 
As before we must also repeat the rigid ideal slab 
calculation using this larger supercell and k-point sampling.
As shown in Table \ref{energy}, the 
energy of the relaxed adatom-reconstructed 
surface turns out to be lower than all previous
reconstructions by as much as 0.09 eV/$(1\times 1)cell.$ 
This indicates that 
the adatom / restatom reconstruction is the most
efficient way of saturating the 
highly unstable dangling bonds of the 
clean (111) surface. The alternative choice 
of a $c(4\times 2)$ adatom / restatom geometry 
(using the two k-points obtained from the refolding in the
$c(4\times 2)$ ISBZ of the 
four ones used in the above $(2\times 1)$ 
reconstruction calculations) confirms this result, 
yielding an energy 16 meV/$(1\times 1)cell$
lower than the ($2\times 2$) choice.
This energy difference is however 
comparable with our overall energy resolution. Similarly, 
our calculation does not rule out the existence of more 
complex surface reconstructions. It does however suggest
that, if this is the case, then the adatom / rest 
atom mechanism, either in a ($2\times 2$) or 
in a $c(4\times 2)$ arrangement, 
is very likely to constitute their basic 
building block, 
as is the case in Si \cite{7x7} and Ge \cite{c2x8}.
Moreover, the advantage over the $\pi$-bonded chain 
reconstruction is larger 
in $\alpha$-Sn, confirming the Si-Ge trend. 

In Table \ref{2x2_pos} 
we report the relaxed atomic coordinates of the 
$(2\times 2)$ adatom-restatom reconstructed surface. 
Due to the symmetry constraint, the adatom and restatom 
only relax in the $z$ direction.
The in-plane distance between the rest atom
and its three neighbors shrinks by about 10 \% with respect 
to the ideal bulk value. 
The rest atom moves outward from its initial
bulk-like position by 0.8 \AA\ , 
again a very large relaxation 
if compared with values in Si (0.3 \AA\ ),
and in Ge (0.55 \AA\ ). Due to this relaxation, the rest atom 
has bond angles of 94.0$^\circ$, a value closer to total $s$, $p$
dehybridization (90$^\circ$) than 
to the original $sp^3$ one (109$^\circ$), 
and bond lengths of 2.84 \AA\
(bulk bond length is 2.79 \AA\ ). 
In turn, the adatom
bond lengths with the three neighboring first-layer atoms 
are 2.97 \AA\ , larger by 7 \% than the bulk bond length, 
and the adatom bond angles are 93.4$^\circ$. 
The second-layer atom under the adatom is pushed
downward. Although its final distance from the adatom (2.99 \AA\ ) 
is comparable with the other adatom bond length, 
there is no bond-like accumulation of electronic charge 
between them, as can be seen in the total charge 
density reported in Fig. \ref{2x2_totcharge}. 
Rather, most of the adatom charge is transferred to 
the restatom dangling bond. 
This charge transfer is not directly visible 
in the total charge density of Fig. \ref{2x2_totcharge}, 
but shows up clearly in the band structure of Fig. \ref{2x2_band}. 
The half filled 
surface-localized band typical of unsaturated dangling bonds 
(see Fig. \ref{ideal_band}), is here split 
into a lower filled band with prevailing rest-atom character,
and an upper empty band manly 
localized onto the adatom. This charge 
transfer is known to be the fundamental mechanism for the 
stabilization of the adatom / rest atom reconstruction in Si and 
Ge \cite{Meade_sige2x2,Takeuchi_ge2x2}, 
and does appear to be so even in Sn. 
A glance at the charge density associated 
with the lower and upper surface bands (see Fig. 
\ref{2x2_charge}) confirms this expectation. 
The charge associated with the lower band, shown 
in the panel (a) of Fig. \ref{2x2_charge}, 
has a strong rest-atom character 
and extends very little into the bulk, 
thus confirming its picture of saturated dangling bond. On the 
contrary, the upper unoccupied 
band (panel (b) of Fig. \ref{2x2_charge})
has a strong adatom character, but 
extends largely below the adatom, 
and might therefore be thought to as 
a band of so called ``floating bonds'' \cite{Pantelides}
associated with both the adatom and the five-fold
coordinated atom immediately underneath. 
   
The optimal atomic positions and 
electronic band structure of the
$c(4\times 2)$ adatom/restatom reconstruction 
are given in Table \ref{c2x4_pos} and Fig. \ref{c2x4_band}. 
Differences in the atomic relaxations and charge densities are
negligible with respect to 
the ($2\times 2$) case, and will not be 
re-discussed.

We have also verified that the $H_3$ site for the adatom is not 
energetically favored, the energy of this surface being 
higher than the one with adatoms in $T_4$ 
sites by about 70 meV/$(1\times 1)cell$. 
Its optimal atomic coordinates are given in Table \ref{H3_pos}.

\section{Metallic overlayers}

The presence of a metallic 
phase ($\beta$-Sn) energetically very close
to the $\alpha$ phase, 
suggests that an insulator-to-metal transition
might take place at the surfaces of $\alpha$-Sn, 
for example, through the formation of
a thin $\beta$-Sn metallic overlayer. Such was 
proposed to be the case, e.g., in gallium\cite{Marco_gallio}, 
where based on calculations similar to these presented here
a metallic bilayer of Ga-III was predicted to
stabilize the surface of $\alpha$-Ga better 
than any other reconstruction.
In our case, however, we encounter a problem, since
no low-index $\beta$-Sn planes appear to match epitaxially the 
$\alpha$-Sn (111) lattice. In particular, we have calculated the
surface energy of several structures 
obtained by covering the $\alpha$-Sn(111)
surface with strained 
epitaxial (100), (110), (111), and (221) planes of
$\beta$-Sn. 
None of these surfaces gave a surface energy comparable,
let alone lower, 
than the ideal (111) surface, all our results being higher with
respect to the $(2\times 2)$ adatom / restatom reconstruction by 
more than 0.45 eV/$(1\times 1) cell$. 
However, this negative 
result cannot of course be taken as a guarantee 
that metallic overlayers 
do not form at the $\alpha$-Sn surface, since 
other more complex configurations, 
beyond our present factory, might 
have to be considered. 
The situation with respect to metallization 
remains therefore open.   

An indirect indication against metallization 
is however provided by surface
energies. One can expect metallization to be favored, 
in fact, in cases
where the metal has a lower 
surface free energy. In our case, however,
the T=0 K energy of 
reconstructed $\alpha$-Sn (111), $E_{surf}\approx
540~mJ/m^2 ,$ is substantially lower 
than the free energy measured for
$\beta$-Sn\cite{Kumikov}, 
namely $670~mJ/m^2 .$ This suggests that 
metallization might not 
take place in the ground state, while it could 
be attained after deconstruction.

\section{Discussion and Conclusions}

From this {\em ab-initio} study, 
we predict that the (111) surface 
of $\alpha$-Sn should be unstable in its unreconstructed form, 
and can be stabilized by various types of reconstructions.

Among the reconstructions not involving adatoms, 
the $(2\times 1)$  $\pi$-bonded
chain model is energetically best, 
and is predicted to display a gigantic
buckling above 1 \AA\ in magnitude. 
This is in line with the trend
towards increasing buckling 
in going from diamond to Si to Ge (111).

In the presence of adatoms, 
the adatom-restatom pair reconstruction is found
to be energetically best, 
and is thus the strongest candidate for the
true ground state of this surface. The relative energy gain over
the $\pi$-bonded state is 
larger than in Si and Ge, again in line with trends. 
The actual optimal periodicity of the adatom-reconstructed
$\alpha$-Sn (111) is however difficult to predict (as is the case
in Si and Ge) because of many subtle factors, including possible
coexistence and competition 
between the two different basic units,
namely $(2\times 2)$ and $c(4\times 2).$

A search for a possible metallic 
state of $\alpha$-Sn (111), suggested by 
the proximity of $\beta$-Sn in the phase diagram, 
was so far inconclusive,
and is disfavored by surface energy considerations.

The beginning of actual experiments on $\alpha$-Sn (111) 
is at this stage 
highly desirable, as we are not aware of any data so far.
The present study makes rather
strong predictions, 
both at the qualitative and the quantitative level.
Qualitatively, we expect $\pi$-bonded chains to dominate on the 
adatom-free (e.g., cleaved) surface, 
and adatom-restatom reconstructions 
to appear after annealing, very much as for Si and Ge (111). 
At the quantitative level, we predict that buckling distortions 
and the associated dipole moments connected
with intra-surface ionic charge transfers should be larger than
in Si and Ge. This effect ought to be relevant to, e.g., surface
vibrational spectroscopy. 
The corresponding band splittings should finally
be observable in electron spectroscopies, 
including photoemission and 
optical absorption, 
particularly to optical conductivity for $q\neq 0.$

It is hoped that these results will stimulate newer efforts
towards an understanding of the $\alpha$-Sn (111) surface, 
where hardly any data can presently be found.

\centerline{\bf  ACKNOWLEDGMENTS}

We would like to thank Dr. Marco Bernasconi 
for providing us the computer
codes used in this work, and for his cooperation in the early
stage of this work. This research was partly sponsored through EEC,
contract ERBCHRXCT 930342, and by the Italian CNR, under projects
``SUPALTEMP'' and also under contract ``95.01056.CT12''.

\newpage

\widetext

\begin{table}
\caption{  }\label{trend}
\begin{tabular}{|c|c|c|c|}
              & C 	&	Si	&	Ge\\ \hline
    $2 \times 1$       
    $\pi$-bonded  
chains  & dimerized ($\ast$) (?) & buckled  & more buckled\\ \hline  
    adatom/
    rest atom    &     not found        & $7 \times 7$ 
  DAS model ($\ast$) &   c$(2 \times 8)$ 
	 adatom model ($\ast$)\\ \hline
   $2 \times 1$
     buckled &      not found       
&  saddle-point &      saddle-point \\ \hline
   metallic  & graphitic $T\ge 2700$K (?)  
&  $T\ge 1500$K (?) & $T\ge 1050$K    \\       
\end{tabular}
\noindent ($\ast$) Stable structure at $T=0$.\\
\noindent (?) Pending confirmation.
\end{table}

\begin{table}
\caption{Calculated surface energies, 
absolute and relative, of different 
optimized reconstructions for $\alpha$-Sn (111) surface.
}\label{energy}
\begin{tabular}{|l|c|c|c|}
Structure & $E_{surf}$ & $\Delta E$ &
vert. relax. or buckling\tablenotemark[1]
 \\
& \begin{tabular}{l|l}
[eV/(1$\times$1 cell)]~~&~~
[$mJ/m^2$] \end{tabular} & [eV/(1$\times$1 cell)]&
( \AA\ ) \\ \hline
ideal             &\begin{tabular}{l|l} 0.940~~~~~~~~~~&~~~~
837 \end{tabular}
 &  0.000  & 0 \\ \hline
fully relaxed     &\begin{tabular}{l|l} 0.933~~~~~~~~~~&~~~~
831 \end{tabular}
 & -0.007  & 0.02\\ \hline
$2\times 1$ buckled (Haneman)
&\begin{tabular}{l|l} 0.732~~~~~~~~~~&~~~~
652 \end{tabular}
 & -0.217  & 1.23  \\ \hline
$2\times 1 ~\pi$-bonded chain&\begin{tabular}{l|l} 
0.697~~~~~~~~~~&~~~~
621 \end{tabular}
 & -0.243  & 1.15  \\ \hline
2$\times$2-adatom ($H_3$)   
 &\begin{tabular}{l|l} 0.696~~~~~~~~~~&~~~~
620 \end{tabular}
 & -0.244  & 
0.54 (restatom)\\  \hline
2$\times$2-adatom ($T_4$) 
&\begin{tabular}{l|l} 0.626~~~~~~~~~~&~~~~
558 \end{tabular}
 & -0.314  & 
0.81 (restatom)\\  \hline
c(4$\times$2)-adatom ($T_4$) 
&\begin{tabular}{l|l} 0.610~~~~~~~~~~&~~~~
543 \end{tabular}
 & -0.330  & 
0.74 (restatom)\\  \hline
metallized   &\begin{tabular}{l|l} $\ge$ 1.06~~~~~~~~&~~~~
944 \end{tabular}
 & $\ge$ 0.121  & \\ 
\end{tabular}
\tablenotemark[1]{Vertical relaxation 
is relative to initial bulk-like 
positions, and buckling is difference 
between relaxation of two atoms.}
\end{table}

\begin{table}
\caption{Ideal, and optimized atomic positions of 
the $\alpha$-Sn (111) ($2\times 1$) 
buckled (Haneman) surface. 
In the rectangular supercell, coordinates 
are given by ${\bf r}=c_1{\bf a}_1+c_2{\bf a}_2+c_3{\bf a}_3,$
where ${\bf a}_i$ is defined 
in the conventional cubic coordinate system
as ${\bf a}_1=(a_0/2)(-1, 2, -1),$ ${\bf a}_2=(a_0/2)(-1, 0, 1),$
${\bf a}_3=a_0(1, 1, 1)$ and 
$a_0$ (=6.446 \AA\ ) is the lattice parameter.}
\label{buck_pos}\begin{tabular}{ccccccc}
Atom& &Ideal& & &Optimal& \\
no. &$c_1$& $c_2$& $c_3$& $c_1$& $c_2$& $c_3$  \\ \hline
  1    &    .000  &    .000  &    .000
       &    .009  &    .000  &   -.048   \\
  2    &    .500  &    .500  &    .000
       &    .485  &    .500  &    .062   \\
  3    &    .167  &    .500  &    .083
       &    .132  &    .500  &    .077   \\
  4    &    .667  &    .000  &    .083
       &    .692  &    .000  &    .077   \\
  5    &    .167  &    .500  &    .333
       &    .169  &    .500  &    .330   \\
  6    &    .667  &    .000  &    .333
       &    .667  &    .000  &    .332   \\
  7    &    .333  &    .000  &    .417
       &    .334  &    .000  &    .415   \\
  8    &    .833  &    .500  &    .417
       &    .836  &    .500  &    .417   \\
  9    &    .333  &    .000  &    .667
       &    .334  &    .000  &    .665   \\
 10    &    .833  &    .500  &    .667
       &    .833  &    .500  &    .667   \\
\end{tabular}
\end{table}

\begin{table}
\caption{ Initial (Pandey's), and optimized 
atomic positions of the $\alpha$-Sn (111) ($2\times 1$)
$\pi$-bonded surface. In the rectangular supercell, coordinates 
are given by ${\bf r}=c_1{\bf a}_1+c_2{\bf a}_2+c_3{\bf a}_3,$
where ${\bf a}_i$ is defined 
in the conventional cubic coordinate system
as ${\bf a}_1=(a_0/2)(-1, 2, -1),$ ${\bf a}_2=(a_0/2)(-1, 0, 1),$
${\bf a}_3=a_0(1, 1, 1)$ and $a_0$ (= 6.446 \AA\ ) is the lattice 
parameter.}\label{pi_pos}
\begin{tabular}{ccccccc}
Atom& &Pandey's\tablenotemark[1]& & &Optimal& \\
no. &$c_1$& $c_2$& $c_3$& $c_1$& $c_2$& $c_3$  \\ \hline
  1    &   -.148  &    .000  &    .023
       &   -.167  &    .000  &    .060   \\
  2    &   -.019  &    .500  &    .023
       &   -.020  &    .500  &   -.043   \\
  3    &    .315  &    .500  &    .106
       &    .278  &    .500  &    .100   \\
  4    &    .519  &    .000  &    .106
       &    .484  &    .000  &    .077   \\
  5    &    .167  &    .500  &    .333
       &    .178  &    .500  &    .341   \\
  6    &    .667  &    .000  &    .333
       &    .657  &    .000  &    .317   \\
  7    &    .333  &    .000  &    .417
       &    .336  &    .000  &    .431   \\
  8    &    .833  &    .500  &    .417
       &    .831  &    .500  &    .400   \\
  9    &    .333  &    .000  &    .667
       &    .337  &    .000  &    .678   \\
 10    &    .833  &    .500  &    .667
       &    .834  &    .500  &    .654   \\
\end{tabular}
\tablenotemark[1]{For corresponding bulk-like positions, see the
ideal case of Table \ref{buck_pos}.}
\end{table}  

\begin{table}
\caption{ Initial, and optimized atomic positions of 
the $\alpha$-Sn (111) ($2\times 2$) adatom / restatom  
($T_4$ configuration) reconstructed surface. 
In the hexagonal supercell, coordinates are given 
by ${\bf r}=c_1{\bf a}_1+c_2{\bf a}_2+c_3{\bf a}_3,$ where 
${\bf a}_i$ is defined 
in the conventional cubic coordinate system
as ${\bf a}_1=(a_0/2)(-1, 2, -1),$ ${\bf a}_2=(a_0/2)(-1, 0, 1),$
${\bf a}_3=(a_0/3)(1, 1, 1)$ and $a_0$ (= 6.446 \AA\ ) is the
lattice parameter.}\label{2x2_pos}
\begin{tabular}{ccccccc}
Atom& &Initial& & &Optimal& \\
no. &$c_1$& $c_2$& $c_3$& $c_1$& $c_2$& $c_3$  \\ \hline
  1    &    .000  &    .000  &   2.375
       &    .000  &    .000  &   2.509   \\
  2    &   -.167  &    .500  &   2.125
       &   -.158  &    .475  &   2.077   \\
  3    &    .333  &    .000  &   2.125
       &    .317  &    .000  &   2.077   \\
  4    &   -.167  &   -.500  &   2.125
       &   -.158  &   -.475  &   2.077   \\
  5    &    .333  &  -1.000  &   2.125
       &    .333  &  -1.000  &   2.343   \\
  6    &    .000  &    .000  &   1.875
       &    .000  &    .000  &   1.705   \\
  7    &    .500  &    .500  &   1.875
       &    .485  &    .544  &   1.935   \\
  8    &    .000  &  -1.000  &   1.875
       &    .030  &  -1.000  &   1.935   \\
  9    &    .500  &   -.500  &   1.875
       &    .485  &   -.544  &   1.935   \\
 10    &    .000  &    .000  &   1.125
       &    .000  &    .000  &    .963   \\
 11    &    .500  &    .500  &   1.125
       &    .500  &    .501  &   1.175   \\
 12    &    .000  &  -1.000  &   1.125
       &    .000  &  -1.000  &   1.175   \\
 13    &    .500  &   -.500  &   1.125
       &    .500  &   -.501  &   1.175   \\
 14    &    .167  &    .500  &    .875
       &    .175  &    .525  &    .864   \\
 15    &    .667  &    .000  &    .875
       &    .667  &    .000  &    .926   \\
 16    &    .167  &   -.500  &    .875
       &    .175  &   -.525  &    .864   \\
 17    &    .667  &  -1.000  &    .875
       &    .650  &  -1.000  &    .864   \\
 18    &    .167  &    .500  &    .125
       &    .170  &    .510  &    .110   \\
 19    &    .667  &    .000  &    .125
       &    .667  &    .000  &    .175   \\
 20    &    .167  &   -.500  &    .125
       &    .170  &   -.510  &    .110   \\
 21    &    .667  &  -1.000  &    .125
       &    .660  &  -1.000  &    .110   \\
\end{tabular}
\end{table}  

\begin{table}
\caption{Initial, and optimized atomic 
positions of the $\alpha$-Sn (111) 
$c(4\times 2)$ adatom / restatom ($T_4$ configuration) 
reconstructed surface. In the rectangular supercell, coordinates 
are given by ${\bf r}=c_1{\bf a}_1+c_2{\bf a}_2+c_3{\bf a}_3,$
where ${\bf a}_i$ is defined 
in the conventional cubic coordinate system
as ${\bf a}_1=(a_0/2)(-1, 2, -1),$ 
${\bf a}_2=(a_0/2)(-1, 0, 1),$
${\bf a}_3=(a_0/3)(1, 1, 1)$ and $a_0$ (=6.446\AA\ )
is the lattice parameter.}\label{c2x4_pos}
\begin{tabular}{ccccccc}
Atom& &Initial& & &Optimal& \\
no. &$c_1$& $c_2$& $c_3$& $c_1$& $c_2$& $c_3$  \\ \hline
  1    &    .000  &    .000  &   2.375
       &    .004  &    .000  &   2.527   \\
  2    &    .167  &    .500  &   2.125
       &    .158  &    .469  &   2.083   \\
  3    &   -.333  &    .000  &   2.125
       &   -.313  &    .000  &   2.102   \\
  4    &    .167  &   -.500  &   2.125
       &    .158  &   -.469  &   2.083   \\
  5    &   -.333  &  -1.000  &   2.125
       &   -.341  &  -1.000  &   2.325   \\
  6    &    .000  &    .000  &   1.875
       &   -.006  &    .000  &   1.716   \\
  7    &   -.500  &    .500  &   1.875
       &   -.489  &    .525  &   1.941   \\
  8    &    .000  &  -1.000  &   1.875
       &   -.027  &  -1.000  &   1.945   \\
  9    &   -.500  &   -.500  &   1.875
       &   -.489  &   -.525  &   1.941   \\
 10    &    .000  &    .000  &   1.125
       &    .001  &    .000  &    .975   \\
 11    &   -.500  &    .500  &   1.125
       &   -.501  &    .501  &   1.178   \\
 12    &    .000  &  -1.000  &   1.125
       &    .005  &  -1.000  &   1.182   \\
 13    &   -.500  &   -.500  &   1.125
       &   -.501  &   -.501  &   1.178   \\
 14    &   -.167  &    .500  &    .875
       &   -.174  &    .525  &    .880   \\
 15    &   -.667  &    .000  &    .875
       &   -.650  &    .000  &    .874   \\
 16    &   -.167  &   -.500  &    .875
       &   -.174  &   -.525  &    .880   \\
 17    &   -.667  &  -1.000  &    .875
       &   -.665  &  -1.000  &    .907   \\
 18    &   -.167  &    .500  &    .125
       &   -.169  &    .509  &    .127   \\
 19    &   -.667  &    .000  &    .125
       &   -.661  &    .000  &    .123   \\
 20    &   -.167  &   -.500  &    .125
       &   -.169  &   -.509  &    .127   \\
 21    &   -.667  &  -1.000  &    .125
       &   -.668  &  -1.000  &    .149   \\
\end{tabular}
\end{table}  

\begin{table}
\caption{Initial, and optimized atomic positions of 
the $\alpha$-Sn(111) ($2\times 2$) adatom / restatom 
($H_3$ configuration) reconstructed surface. 
In the hexagonal supercell, coordinates are given 
by ${\bf r}=c_1{\bf a}_1+c_2{\bf a}_2+c_3{\bf a}_3,$
where ${\bf a}_i$ is defined 
in the conventional cubic coordinate system
as ${\bf a}_1=(a_0/2)(-1, 2, -1),$ 
${\bf a}_2=(a_0/2)(-1, 0, 1),$
${\bf a}_3=(a_0/3)(1, 1, 1)$ and $a_0$ (= 6.446 \AA\ ) is the 
lattice parameter.} \label{H3_pos}
\begin{tabular}{ccccccc}
Atom& &Initial& & &Optimal& \\
no. &$c_1$& $c_2$& $c_3$& $c_1$& $c_2$& $c_3$  \\ \hline
  1    &    .167  &    .500  &   2.375
       &    .167  &    .500  &   2.525   \\
  2    &   -.167  &    .500  &   2.125
       &   -.142  &    .500  &   2.090   \\
  3    &    .333  &    .000  &   2.125
       &    .321  &    .037  &   2.090   \\
  4    &   -.167  &   -.500  &   2.125
       &   -.167  &   -.500  &   2.271   \\
  5    &    .333  &  -1.000  &   2.125
       &    .321  &  -1.037  &   2.090   \\
  6    &    .000  &    .000  &   1.875
       &   -.015  &   -.046  &   1.848   \\
  7    &    .500  &    .500  &   1.875
       &    .531  &    .500  &   1.848   \\
  8    &    .000  &  -1.000  &   1.875
       &   -.015  &   -.954  &   1.848   \\
  9    &    .500  &   -.500  &   1.875
       &    .500  &   -.500  &   1.984   \\
 10    &    .000  &    .000  &   1.125
       &    .000  &   -.001  &   1.102   \\
 11    &    .500  &    .500  &   1.125
       &    .501  &    .500  &   1.102   \\
 12    &    .000  &  -1.000  &   1.125
       &    .000  &   -.999  &   1.102   \\
 13    &    .500  &   -.500  &   1.125
       &    .500  &   -.500  &   1.199   \\
 14    &    .167  &    .500  &    .875
       &    .167  &    .500  &    .861   \\
 15    &    .667  &    .000  &    .875
       &    .663  &   -.012  &    .883   \\
 16    &    .167  &   -.500  &    .875
       &    .175  &   -.500  &    .883   \\
 17    &    .667  &  -1.000  &    .875
       &    .663  &   -.988  &    .883   \\
 18    &    .167  &    .500  &    .125
       &    .167  &    .500  &    .105   \\
 19    &    .667  &    .000  &    .125
       &    .665  &   -.006  &    .133   \\
 20    &    .167  &   -.500  &    .125
       &    .171  &   -.500  &    .133   \\
 21    &    .667  &  -1.000  &    .125
       &    .665  &   -.994  &    .133   \\
\end{tabular}
\end{table}  

\figure{
\caption{Electronic structure of bulk $\alpha$-Sn. 
The zero in energy corresponds to the Fermi level. }
\label{bulk_band}
}

\figure{
\caption{Brillouin zone and 
electronic structure of bulk $\beta$-Sn.
The zero in energy corresponds to the Fermi level. }
\label{beta_band}
}

\figure{
\caption{Surface electronic structure 
of the ideal $\alpha$-Sn(111) surface
reported along high-symmetry lines of 
the $(1\times 1)$ hexagonal irreducible Brillouin zone.
Shaded areas correspond to surface-projected bulk states, 
while thicker lines correspond to surface states. 
The surface Brillouin zone is given in the inset. Note that the 
dangling bond surface state crosses the fermi level $E_F$.}
\label{ideal_band}
}

\figure{
\caption{ Electron density contour of the highest occupied 
state of the ideal $(1\times 1)$ surface at $\bar{K}$, 
on the (110) plane passing through
top atoms. Full circles correspond to Sn atoms, and thicker
straight lines to bonds among Sn atoms. 
Contour lines are separated by 0.0005 (a.u.). 
Note the dangling bond character of this state. }    
\label{ideal_charge}
}

\figure{
\caption{ Electron density contours of 
the highest occupied state ( panel (a) ) and of the 
lowest unoccupied state ( panel (b) ) in the $(2\times 1)$
$\pi$-bonded chain reconstruction at $\bar{J}$, 
on the (110) plane passing
through the up (down) atom. In panel 
(b) atoms are labelled according to the optimal positions of
Table \protect\ref{pi_pos}. 
Contour lines are separated by 0.0005 (a.u.).}
\label{2x1_charge}
}

\figure{
\caption{Surface electronic structure 
of the $(2\times 1)$ $\pi$-bonded chain
reconstructed surface 
reported in the $(2\times 1)$ rectangular irreducible
Brillouin zone (see inset). Shaded areas correspond 
to surface-projected bulk states, while thicker 
lines correspond to surface states. 
The original dangling bond is now split into
two.} 
\label{2x1_band}
}

\figure{
\caption{Geometry, and electron density contour of the 
$(2\times 2)$ adatom / restatom reconstruction, on the 
plane passing through the adatom and the rest atom. 
Contour lines are separated by 0.005 (a.u.). Note the 
strong outward relaxation of the restatom, 
and the strong inward relaxation of the second 
and third layer atoms 
beneath the adatom. (R: restatom, A: adatom)}
\label{2x2_totcharge}
}

\figure{
\caption{Surface electronic structure 
of the $(2\times 2)$ adatom / restatom
reconstructed surface reported 
in the $(2\times 2)$ hexagonal irreducible
Brillouin zone (see inset). Shaded areas correspond 
to surface-projected bulk states, thicker 
lines correspond to surface states, 
dotted lines to surface resonances.
The $\sim 1$ eV splitting of the surface 
state reflects the adatom-restatom
electron transfer.}
\label{2x2_band}
}

\figure{
\caption{ Electron density contours of 
the highest occupied state ( panel (a) ) and of the 
lowest unoccupied state ( panel (b) ) in the $(2\times 2)$
adatom / restatom  reconstruction at $\bar{K}$, 
on the same plane as
in Fig. \protect\ref{2x2_totcharge}. 
In panel (b) atoms are labelled according to the 
optimal positions of Table \protect\ref{2x2_pos}. 
Contour lines are separated by 0.0004 (a.u.). Note the 
strong restatom / adatom characters, 
with a larger penetration of the
adatom empty state. }
\label{2x2_charge} 
}

\figure{
\caption{Surface electronic structure 
of the $c(4\times 2)$ adatom / restatom
reconstructed surface reported 
in the $c(4\times 2)$ rectangular irreducible
Brillouin zone (see inset). Shaded areas correspond 
to surface-projected bulk states, thicker 
lines correspond to surface states, 
dotted lines to surface resonances. }
\label{c2x4_band}
}

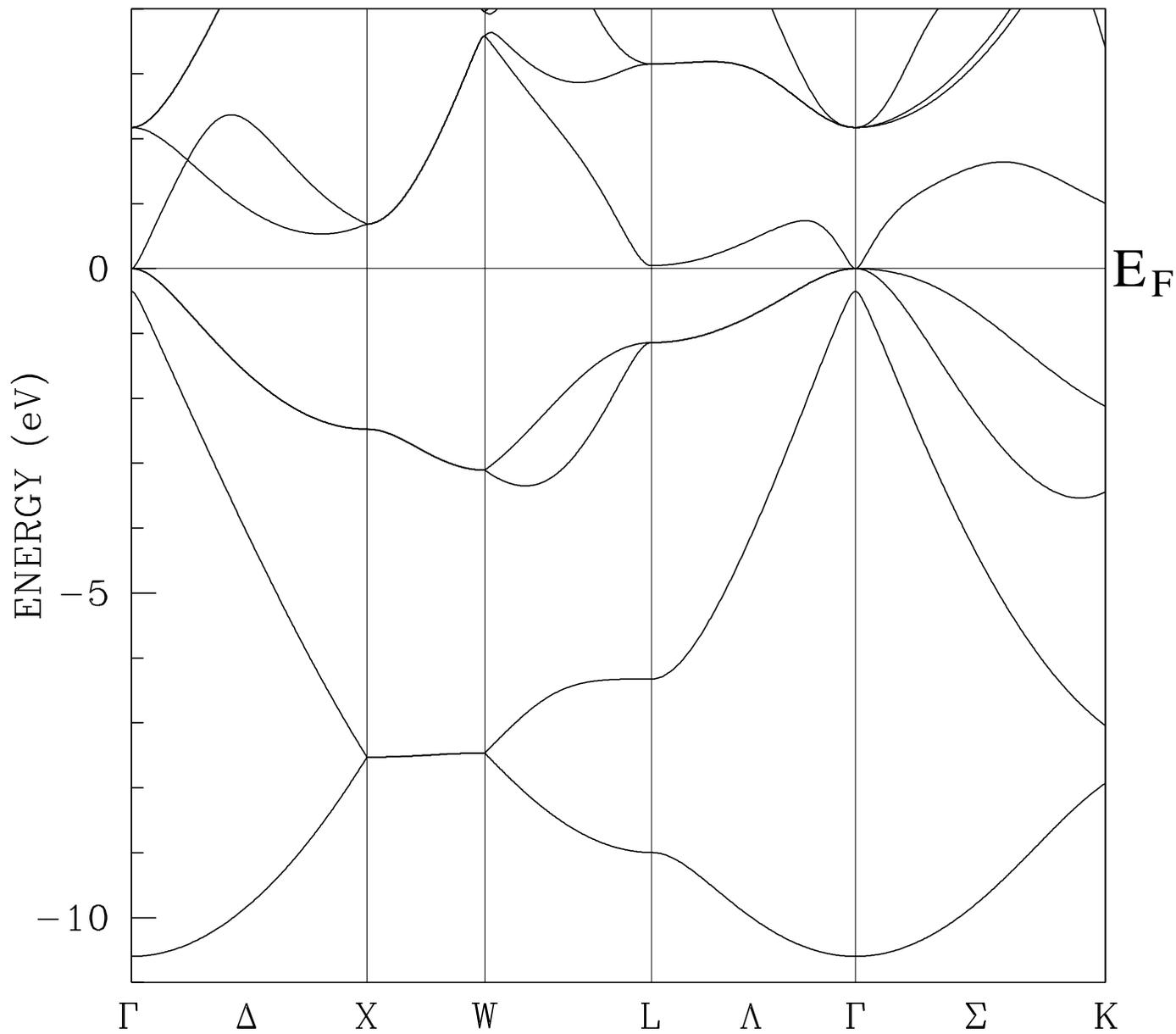
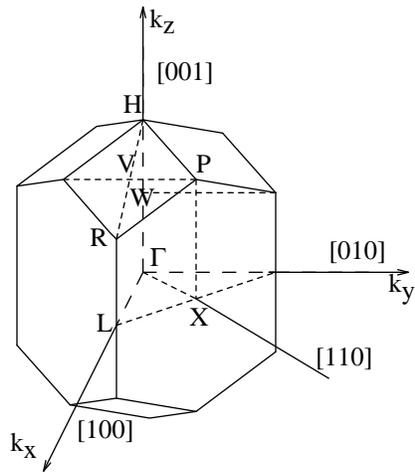
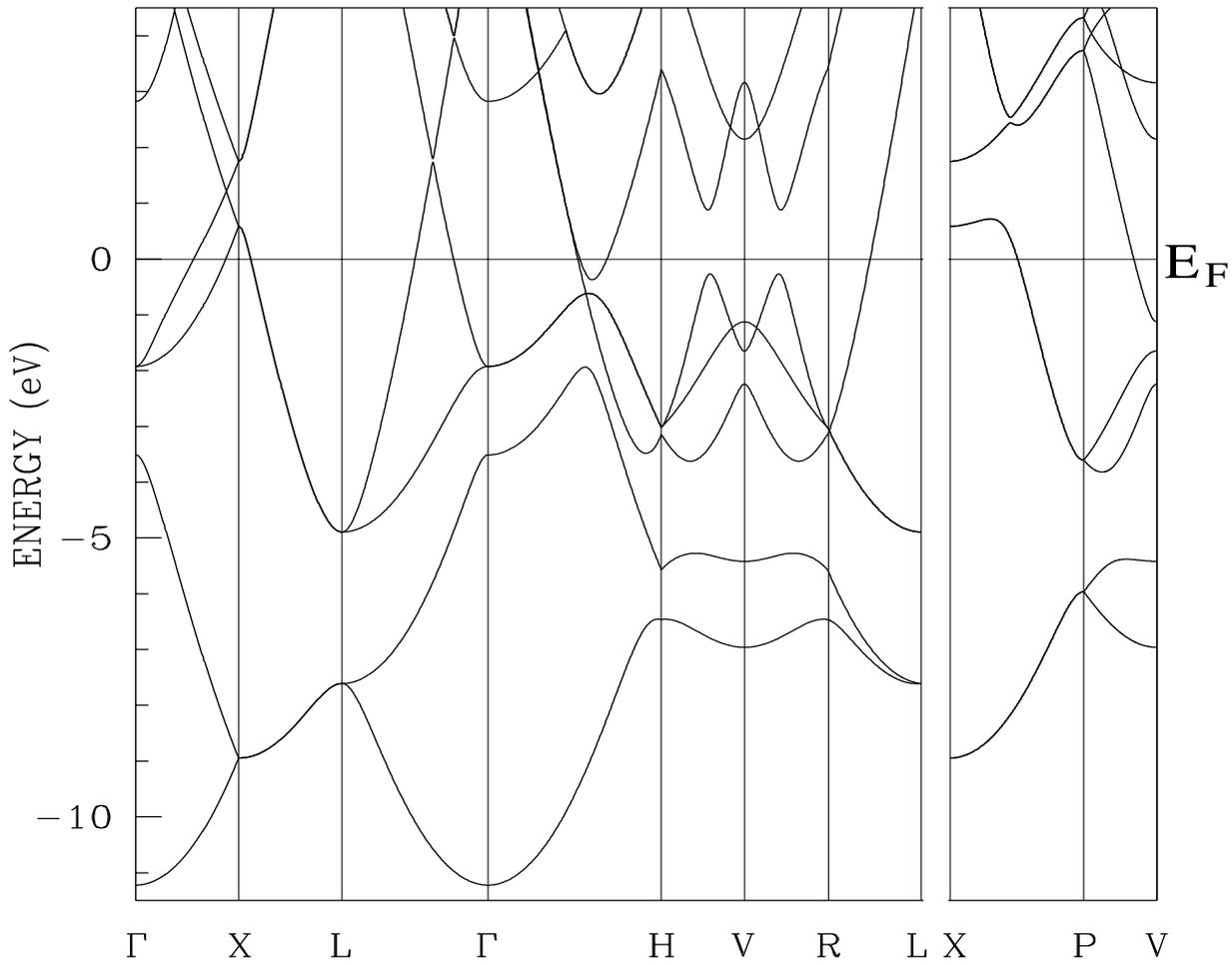
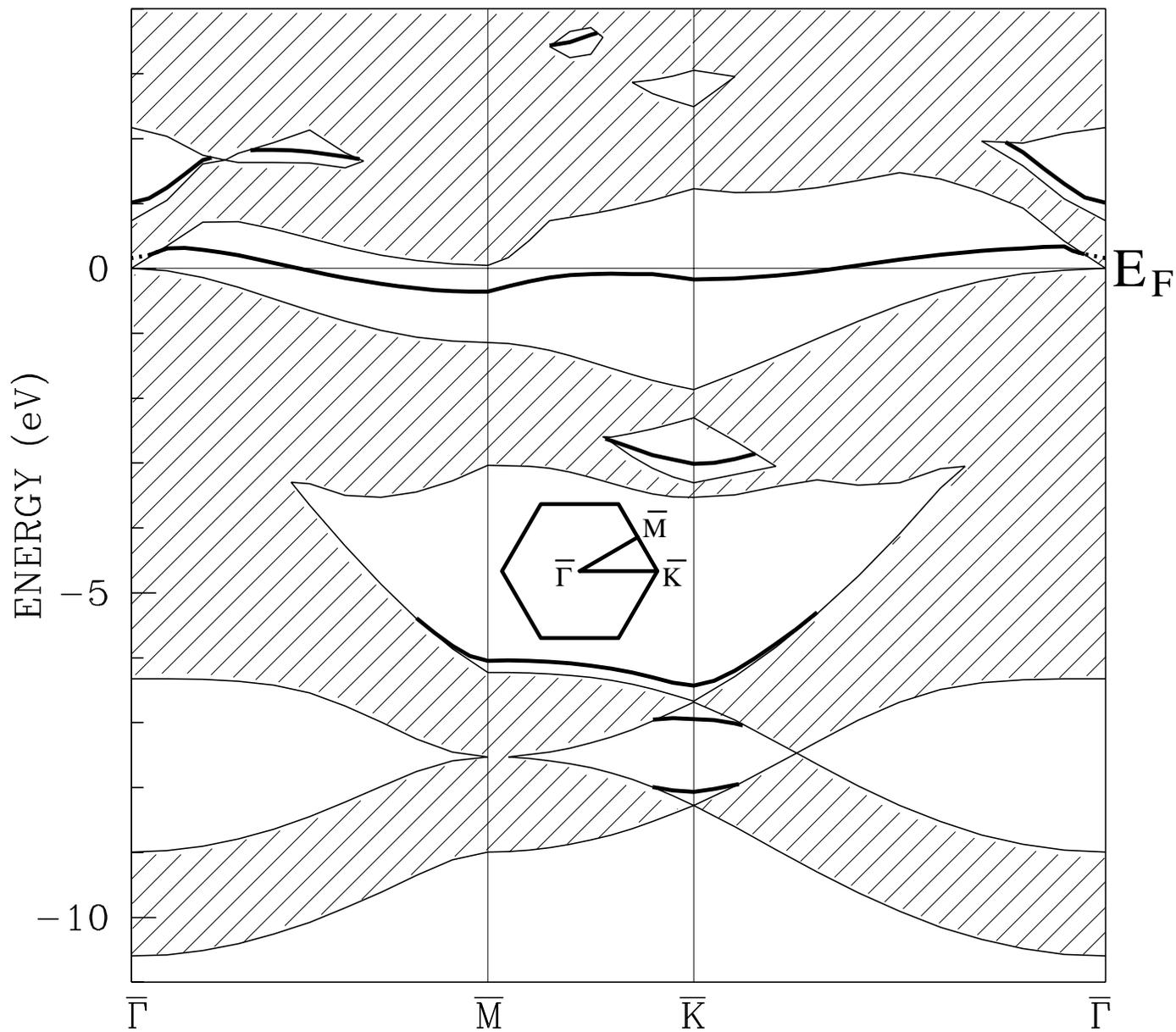
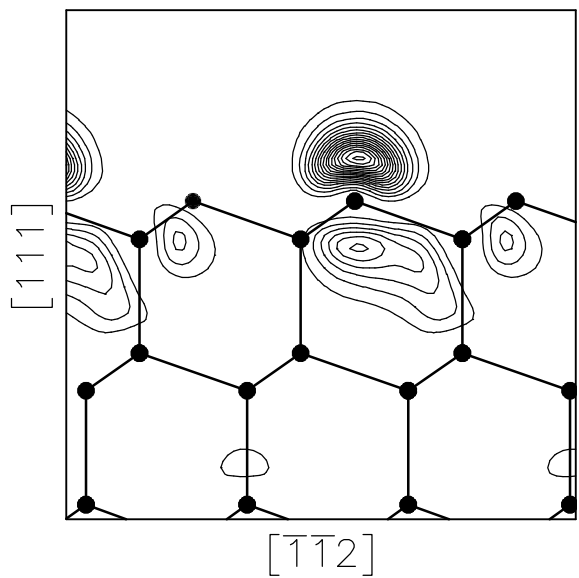
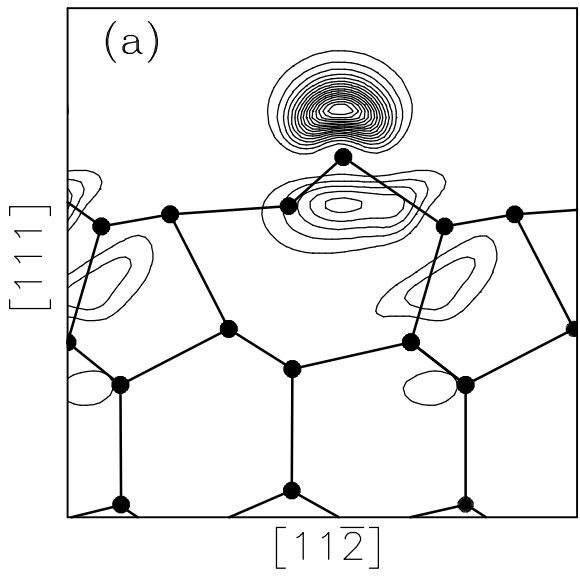
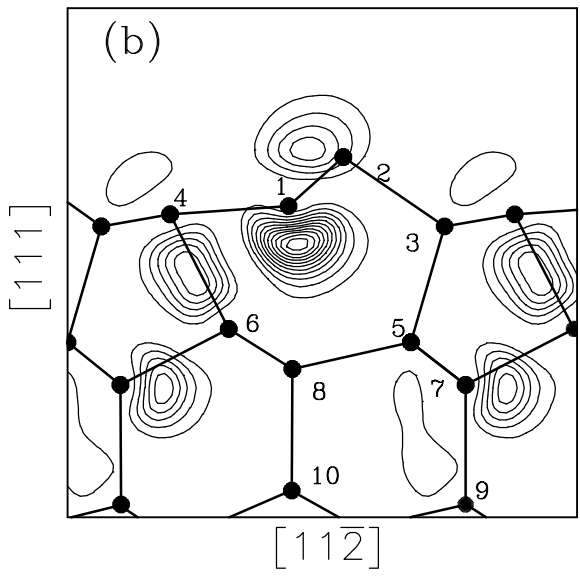
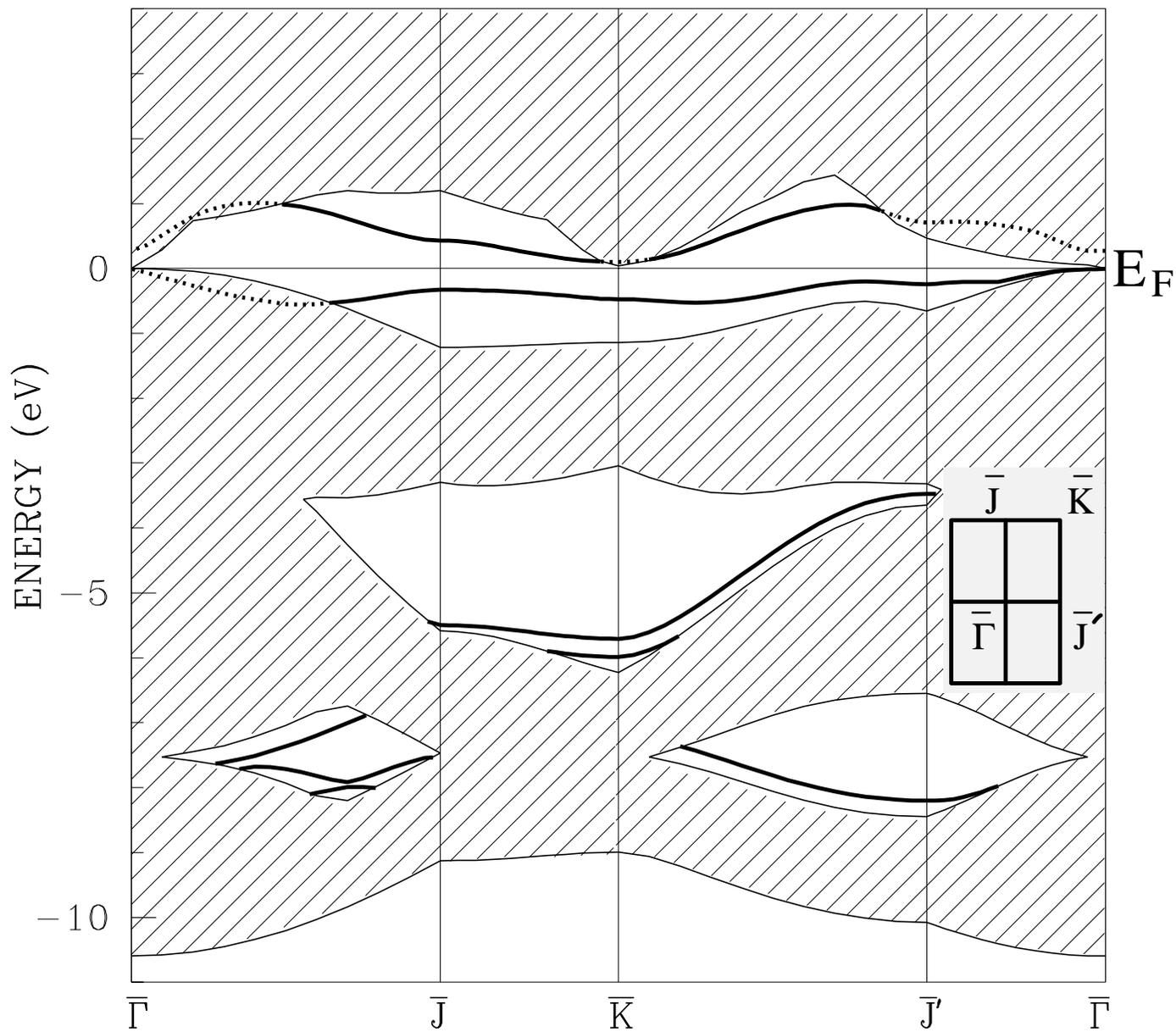
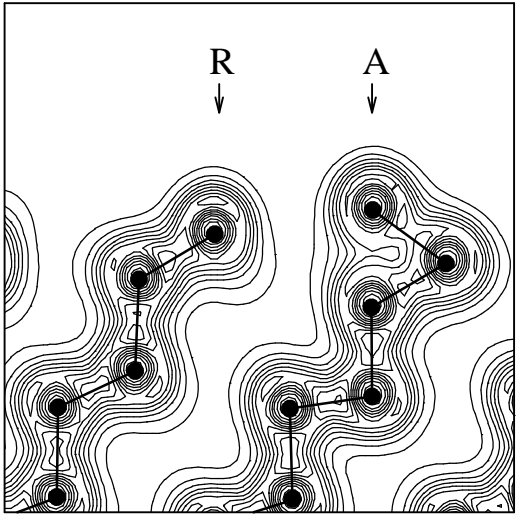
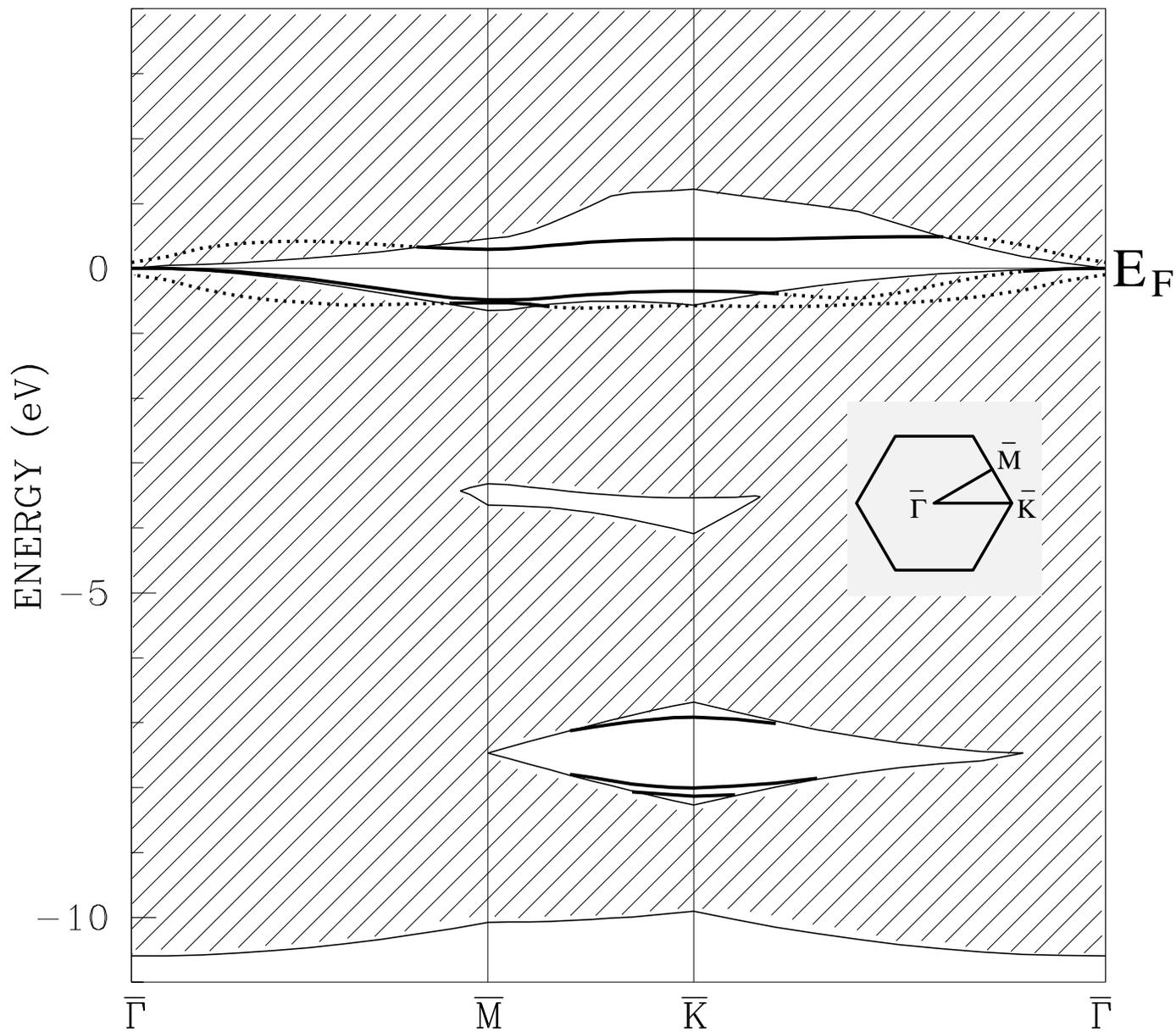
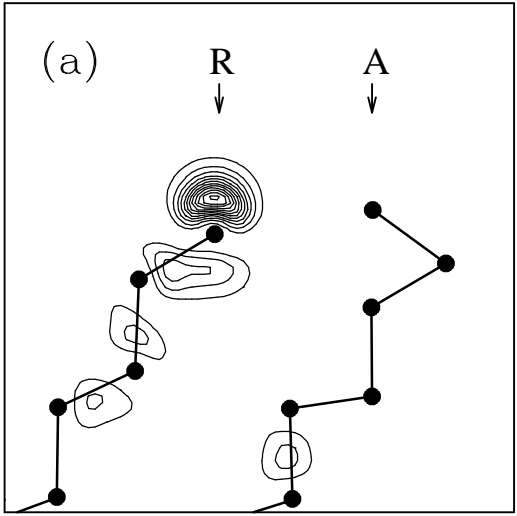
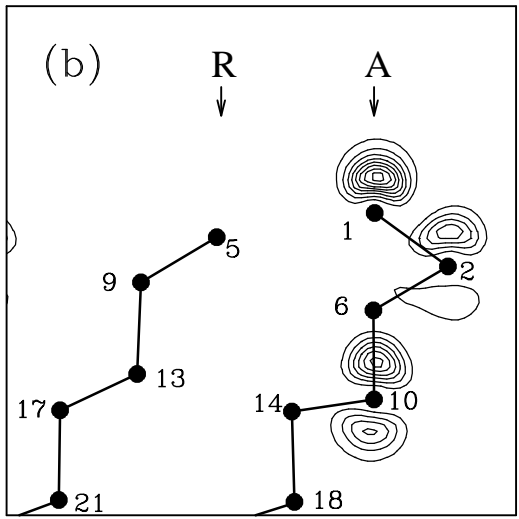
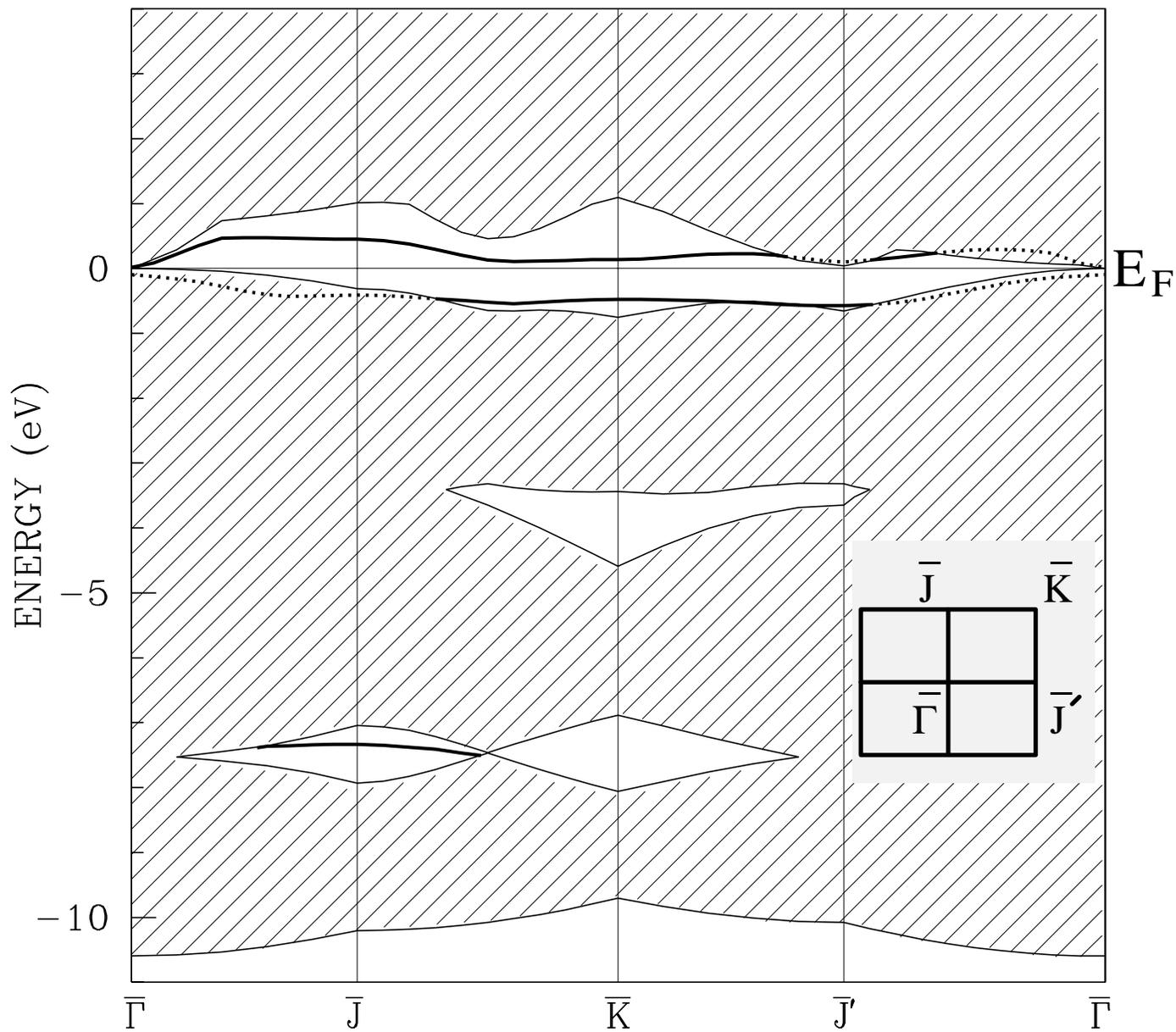
\end{document}